\documentclass[preprints,article,accept,moreauthors,pdftex]{Definitions/mdpi}
\firstpage{1} 
\pubvolume{1}
\issuenum{1}
\articlenumber{0}
\pubyear{2021}
\copyrightyear{2020}
\datereceived{} 
\dateaccepted{} 
\datepublished{} 
\hreflink{https://doi.org/} 

\pdfoutput=1

\Title{Searching for  $H^\pm\to W^{\pm} + 4\gamma$ signals in the 2HDM Type-I at the LHC}

\TitleCitation{Searching for  $H^\pm\to W^{\pm} + 4\gamma$ signals in the 2HDM Type-I at the LHC}

\Author{Yan Wang $^{1,\dagger\ddagger}$\orcidA{}, A. Arhrib $^{2}$, R. Benbrik $^{3,}$, M. Krab $^{4}$, B. Manaut $^{4}$, S. Moretti $^{5}$ and Qi-Shu Yan $^{6}$ $^{7}$}

\address{%
$^{1}$ \quad College of Physics and Electronic Information, Inner Mongolia Normal University, Hohhot 010022, PR China; wangyan@imnu.edu.cn\\
$^{2}$ \quad Abdelmalek Essaadi University, Faculty of Sciences and Techniques,  B.P. 2117 T{\'e}touan, Tanger, Morocco; aarhrib@uae.ac.ma\\
$^{3}$ \quad Laboratoire de Physique Fondamentale et Appliqu{\'e}e de Safi, Facult{\'e} Polydisciplinaire de Safi, Sidi Bouzid, B.P. 4162, Safi, Morocco; r.benbrik@uca.ma\\
$^{4}$ \quad Research Laboratory in Physics and Engineering Sciences, Modern and Applied Physics Team, Polidisciplinary Faculty, Beni Mellal, 23000, Morocco; mohamed.krab@usms.ac.ma, b.manaut@usms.ma\\
$^{5}$ \quad School of Physics and Astronomy, University of Southampton, Southampton SO17 1BJ,
United Kingdom; s.moretti@soton.ac.uk\\
$^{6}$ \quad Center for Future High Energy Physics, Chinese Academy of Sciences, Beijing 100049, PR 
China; \\
$^{7}$ \quad School of Physics Sciences, University of Chinese Academy of Sciences, Beijing 100039,
PR China; yanqishu@ucas.ac.cn\\
}

\corres{Correspondence: wangyan@imnu.edu.cn}

\abstract{    We analyse a new signature of a charged Higgs boson in the Type-I realisation of the 2-Higgs Doublet Model (2HDM), where the mass of the charged Higgs boson satisfies the condition $M_{H^{\pm}}<M_{t}+M_{b}$
and the theoretical parameter space is consistent with the latest experimental constraints from 
 Large Hadron Collider (LHC) and other direct as well as indirect searches. In the surviving regions, it is found that the bosonic decay mode of a charged
Higgs boson is dominated by $H^{\pm} \rightarrow W^{\pm*}h$. At the same time, the light neutral Higgs boson $h$ dominantly
decays into two photons. We thus find that the production process  $pp \rightarrow H^{\pm}h \rightarrow W^{\pm*}h h \rightarrow l^{\pm}\nu +4\gamma$ is
almost background free. Since the $W^{\pm}$ could be either off-shell or on-shell, furthering a previous phenomenological analysis, 
we examine closely detector effects onto the above signature and 
demonstrate that it is indeed promising to study it at the LHC with both $\sqrt{s}= 13$ TeV
and $\sqrt{s} = 14$ TeV,  assuming an integrated luminosity of 300 $fb^{-1}$.}


\newcommand{\beq}{\begin {equation}}
\newcommand{\eeq}{\end   {equation}}
\newcommand{\bea}{\begin {eqnarray}}
\newcommand{\eea}{\end   {eqnarray}}
\newcommand{\baa}{\begin {array}   }
\newcommand{\eaa}{\end   {array}   }
\newcommand{\bit}{\begin {itemize} }
\newcommand{\eit}{\end   {itemize} }
\newcommand{\be }{\begin {equation}}
\newcommand{\ee }{\end   {equation}}

\newcommand{\wboson}{W^{\pm}}

\newcommand{\fbinv}{\mathrm{fb}^{-1}}

\newcommand{\chboson}{H^{\pm}}
\begin{document}
\section{Introduction}
One of the tasks at the  Large Hadron Collider (LHC) is to discover some new physics. In models of it, one or more charged Higgs bosons can be predicted. For example, in the 2-Higgs Doublet Model (2HDM), there are two complex Higgs doublets. After Electro-Weak Symmetry Breaking (EWSB), there are five physical Higgs bosons, 2 neutral CP-even scalars ($h$ and $H$, with $m_h < m_H$, where $m_{h/H}$ is the light(heavy) Higgs boson mass), a CP-odd pseudoscalar ($A$) and two charged Higgs states ($H^{\pm}$). Thus, if a charged Higgs boson can be found, it will be a clear evidence of such a new physics.

In the 2HDM, a $Z_2$ symmetry is introduced into the Yukawa sector in order to avoid too large effects of Flavour Changing Neutral Currents (FCNCs). Depending on the $Z_2$ charge assignment of the Higgs doublets and the fermion flavours, four basic scenarios are defined, which are known as Types-I/II/X/Y of the 2HDM~\cite{Branco:2011iw}. 

In this note, we focus on  $H^\pm\to W^\pm + 4\gamma$ signals in the 2HDM Type-I possibly emerging at the LHC from $H^\pm h$ production and decay, wherein both $h\to\gamma\gamma$~\cite{Wang:2021pxc}. We assume that the heavy scalar $H$ is the observed CP-even SM-like Higgs boson, which properties are consistent with the LHC measurements, and $h$ is lighter than 125 GeV. We  study the properties of a light charged Higgs boson, i.e., satisfying the condition  $M_{\chboson} < M_{t} + M_{b}$. The decay modes of this charged Higgs boson are dominated by $H^{\pm} \to W^{\pm} h$, while the dominant decay mode of the light neutral Higgs boson $h$ is indeed $h \to \gamma \gamma$. This emerges over 
 the parameter space where $h$ is almost  fermiophobic, i.e., $cos\alpha/sin\beta$ $\to$ $0$. In the Type-I scenario, the $hq\bar{q}$ coupling is proportional to $cos\alpha/sin\beta$. Since $cos\alpha=sin\beta sin(\beta-\alpha)+cos\beta cos(\beta-\alpha)$, when $sin(\beta-\alpha)$ is negative and $cos(\beta-\alpha)$ is positive, $cos\alpha$ will be cancelled for a special $tan\beta$. Thus, in the corresponding (fermiophobic) limit,  $h$ decay modes into SM fermions can be highly suppressed (so that, as intimated, $h\to\gamma\gamma$ becomes dominant). Specifically,  we focus on the process $pp \rightarrow H^{\pm}h \rightarrow W^{\pm*}hh \rightarrow l^{\pm}\nu +4\gamma$, i.e., on the aforementioned the $W+4\gamma$ final state. According to the parton level analysis in \cite{Arhrib:2017wmo}, this signature is essentially  background free and can lead to a sizable significance when the integrated luminosity is large enough. In this note, we aim at confirming this result, obtained at the parton level, following a thorough detector level simulation.

We start by performing a scan over the parameter space of the Type-I scenario aimed at maximising the yield of the discussed $W+4\gamma$ signature. Such a scan is based on the latest constraints from both theoretical consistency and experimental bounds by using 2HDMC~\cite{Eriksson:2009ws}, HiggsSignals-2.6.0~\cite{Bechtle:2020uwn} and HiggsBounds-5.9.0~\cite{Bechtle:2020pkv}. 
For the light $h$ state, we vary $m_{h}$ from 20 GeV to about 80 GeV, which is always lighter than 125 GeV and lighter than the $H^{\pm}$ mass too. 
The mass of the charged Higgs boson varies from 91 GeV to 170 GeV. By requiring $H^{\pm} \to W^{\pm}h$, the $W^{\pm}$ still  be on-shell or off-shell. If $M_{H^{\pm}} < M_{W^{\pm}} + M_{h}$, the charged Higgs boson will decay to a soft lepton through an off-shell $W^{\pm}$ boson. Also notice that 
 the parameter $\sin(\beta-\alpha)$ is constrained by the SM-like Higgs boson measurements at the LHC, requiring $-0.3<sin(\beta-\alpha)<0$, and $tan\beta$ is in the range [7,20]. 

\begin{figure}[ht]
\begin{minipage}{0.4\textwidth}
      \begin{center} 
         \includegraphics[height=4cm]{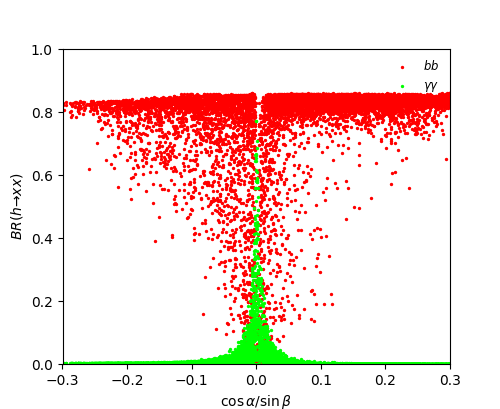}	
      \end{center}
\end{minipage}
\begin{minipage}{0.4\textwidth}
      \begin{center} 
         \includegraphics[height=4cm]{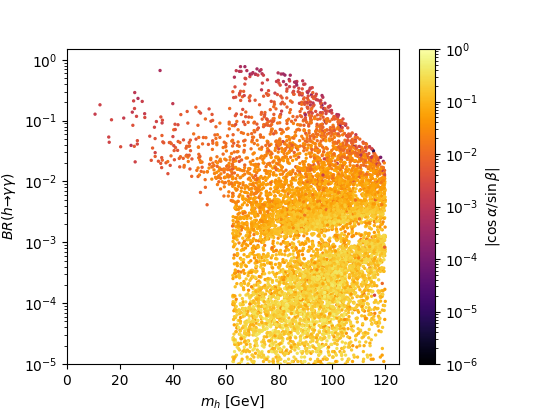}	
      \end{center}
\end{minipage}
    \caption{Scatter plots in $m_h$ and $cos\alpha/sin\beta$ for $BR(h\rightarrow\gamma\gamma, b\bar b)$ are shown.}\label{f:mh_br}
\end{figure}

\begin{figure}[ht]
\begin{minipage}{0.4\textwidth}
      \begin{center} 
         \includegraphics[height=4cm]{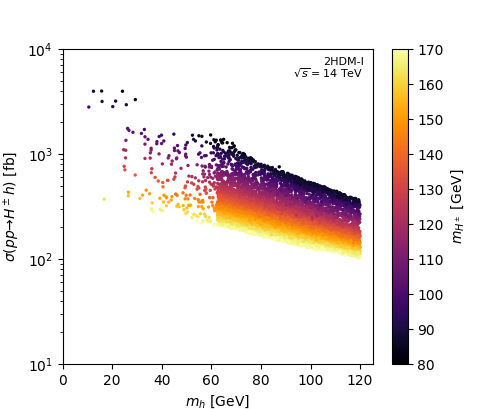}	
      \end{center}
\end{minipage}
\begin{minipage}{0.4\textwidth}
      \begin{center} 
         \includegraphics[height=4cm]{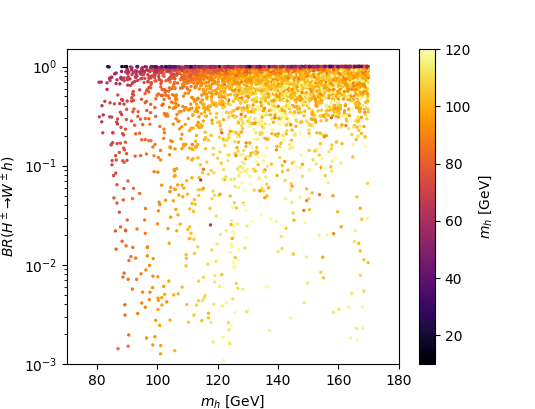}	
      \end{center}
\end{minipage}
    \caption{Scatter plots in $m_h$ and $m_{H^{\pm}}$ for $pp\rightarrow H^{\pm}h$ and $BR(H^{\pm} \rightarrow W^{\pm}h)$ are shown.}\label{f:mh_cx}
\end{figure}

Our results show that the cross section $\sigma$ of our signal process in these regions can reach its maximum value. In Fig.~\ref{f:mh_br}, we show the scatter plots in the plane of $cos\alpha/sin\beta$ (left) or $m_{h}$ (right, where the correlation to $cos\alpha/sin\beta$ is gauged in colour) and Branching Ratios ($BR$s) of the $h$ state, limitedly to the $b\bar b$ and $\gamma\gamma$ channels.
In Fig.~\ref{f:mh_cx}, in the left (right) panel, we show scatter plots to demonstrate the dependence of  $\sigma$($pp\rightarrow H^{\pm}h$) ($BR(H^{\pm} \rightarrow W^{\pm}h$)) on the parameters $m_h$ and $m_{H^\pm}$. From these plots, we selected 14 BPs, which are listed in Tab.~\ref{t_bp_para}. We also deliberately choose $m_h$ and $m_{H^{\pm}}$ to cover the parameter space as much as possible. Most of the BPs lie over the region $m_{h}>62.5$ GeV, which is half of the SM-like Higgs boson mass. It is because, when considering the experiments constraints from  HiggsBounds and HiggsSignals, lots of parameter space will be ruled out for $m_{h}<62.5$ GeV. For BP4--BP10, we implicitly require the $W^{\pm}$ boson to be off-shell.

\begin{table}
\begin{center}
\begin{tabular}{| c| c| c| c| c| c| c| c| c|}
\hline
para&$M_h$&$M_A$&$M_{H^{\pm}}$&$sin(\beta-\alpha)$&$tan\beta$&$M_{12}^{2}$&$\sigma_{13}$ [fb]&$\sigma_{14}$ [fb]\\
\hline
BP1&25.57&72.39&111.08&-0.074&13.58&11.97&101.40&112.55\\
\hline
BP2&35.12&111.24&151.44&-0.075&13.32&16.66&167.75&186.20\\
\hline
BP3&45.34&162.07&128.00&-0.136&7.57&80.96&10.76&11.93\\
\hline
BP4&53.59&126.09&91.49&-0.127&8.00&51.16&27.05&29.88\\
\hline
BP5&63.13&85.59&104.99&-0.056&18.09&190.24&179.31&198.61\\
\hline
BP6&65.43&111.43&142.15&-0.087&11.52&325.36&174.49&194.30\\
\hline
BP7&67.82&79.83&114.09&-0.111&8.94&326.32&177.72&197.23\\
\hline
BP8&69.64&195.73&97.43&-0.111&8.86&357.10&196.04&217.18\\
\hline
BP9&73.18&108.69&97.34&-0.122&8.06&594.64&193.56&214.57\\
\hline
BP10&84.18&115.26&148.09&-0.067&14.82&473.88&61.92&68.98\\
\hline
BP11&68.96&200.84&155.40&-0.112&8.64&531.46&62.02&69.14\\
\hline
BP12&71.99&91.30&160.10&-0.104&9.74&472.22&58.99&65.80\\
\hline
BP13&74.09&102.49&163.95&-0.092&10.56&503.74&55.58&62.04\\
\hline
BP14&81.53&225.76&168.69&-0.101&9.75&501.29&51.85&57.91\\
\hline
\end{tabular}
\end{center}
\caption{Input parameters and parton level cross sections (in fb) corresponding to the selected BPs are tabulated. All masses are in GeV
and for all points $M_H$ = 125 GeV. Here, $\sigma_{13/14}$ denotes the cross section of $pp\to W+4\gamma$ at $\sqrt{s}=13/14$ TeV.}\label{t_bp_para} 
\end{table}

Then, we present a detailed Monte Carlo (MC) analysis at the detector level to examine the feasibility of signal events for the center-of-mass energies of 13 TeV and 14 TeV at the LHC.  The SM background processes include  $W^\pm +4j0\gamma$, $W^\pm+3j1\gamma$, $W^\pm+2j2\gamma$, $W^\pm+1j3\gamma$ and $W^\pm+0j4\gamma$, where one or more jets have a certain probability to fake a photon. 

The MC events are generated by MadGraph5$\_{\rm aMC@NLO}$-2.8.2~\cite{Alwall:2014hca} with the following parton level cuts:
\begin{eqnarray}
|\eta(l,j,\gamma)|<2.5, \quad p_T(j,\gamma,l)>10 ~\text{GeV}, \quad \Delta R(l,j,\gamma)>0.5,  \quad \textrm{MET} > 5 ~\text{GeV}.
\end{eqnarray}
The hadronisation and detector simulation are performed by using Pythia-8~\cite{Sjostrand:2006za} and Delphes-3.4.2~\cite{deFavereau:2013fsa}, where the anti-$k_t$ jet algorithm with jet parameter $\Delta R=0.5$ is adopted in  FastJet~\cite{Cacciari:2011ma} and the fake photon rate for a jet is taken as $0.001$.

After the event pre-selection, it is found that the background events are indeed negligible when compared with the possible signal events for our BPs. Thus, the significance for the  $W+4\gamma$ signature only depends on the signal cross section and the integrated luminosity ${L}$, which can be computed by using the relation $\frac{N_S}{\sqrt{N_S+N_B}}\sim \sqrt{N_S} \sim \sqrt{\sigma \times {L}}$. The final significances for each BP are listed in Tab.~\ref{t_significance}. 
\begin{table}
 \begin{center}
 \begin{tabular}{|c | c| c| c| c| c| c| c| }
 \hline
 BPs & 1 &   2 & 3 &   4  &  5 & 6 & 7    \\
  \hline
 $\sigma_{13 \text{TeV}}$ &12.1& 23.7& 6.7 & 9.4 & 27.4 & 32.6 & 29.2 \\
   \hline
 $\sigma_{14 \text{TeV}}$& 12.5& 24.4 &7.0 & 9.8 & 28.4 & 33.9 & 30.3 \\
  \hline
 BPs &  8 & 9 & 10 & 11 & 12 & 13 & 14   \\
  \hline
 $\sigma_{13 \text{TeV}}$ & 25.2 & 23.9 & 20.8 & 20.2 & 20.3 & 19.9 & 19.9 \\
   \hline
 $\sigma_{14 \text{TeV}}$&  26.2 & 24.8 & 21.8 & 21.1 & 21.0 & 20.8 & 20.8 \\
  \hline
 \end{tabular}
 \end{center}
      \caption{The significances for all 14 BPs at the LHC are tabulated, where the luminosity is assumed to be 300 fb$^{-1}$ at both $\sqrt{s} =13$ and $14$ TeV. } \label{t_significance}
\end{table}

\begin{figure}[!t]
\begin{center}
\end{center}
\begin{minipage}{0.4\textwidth}
    \begin{center}    
 \includegraphics[height=4cm]{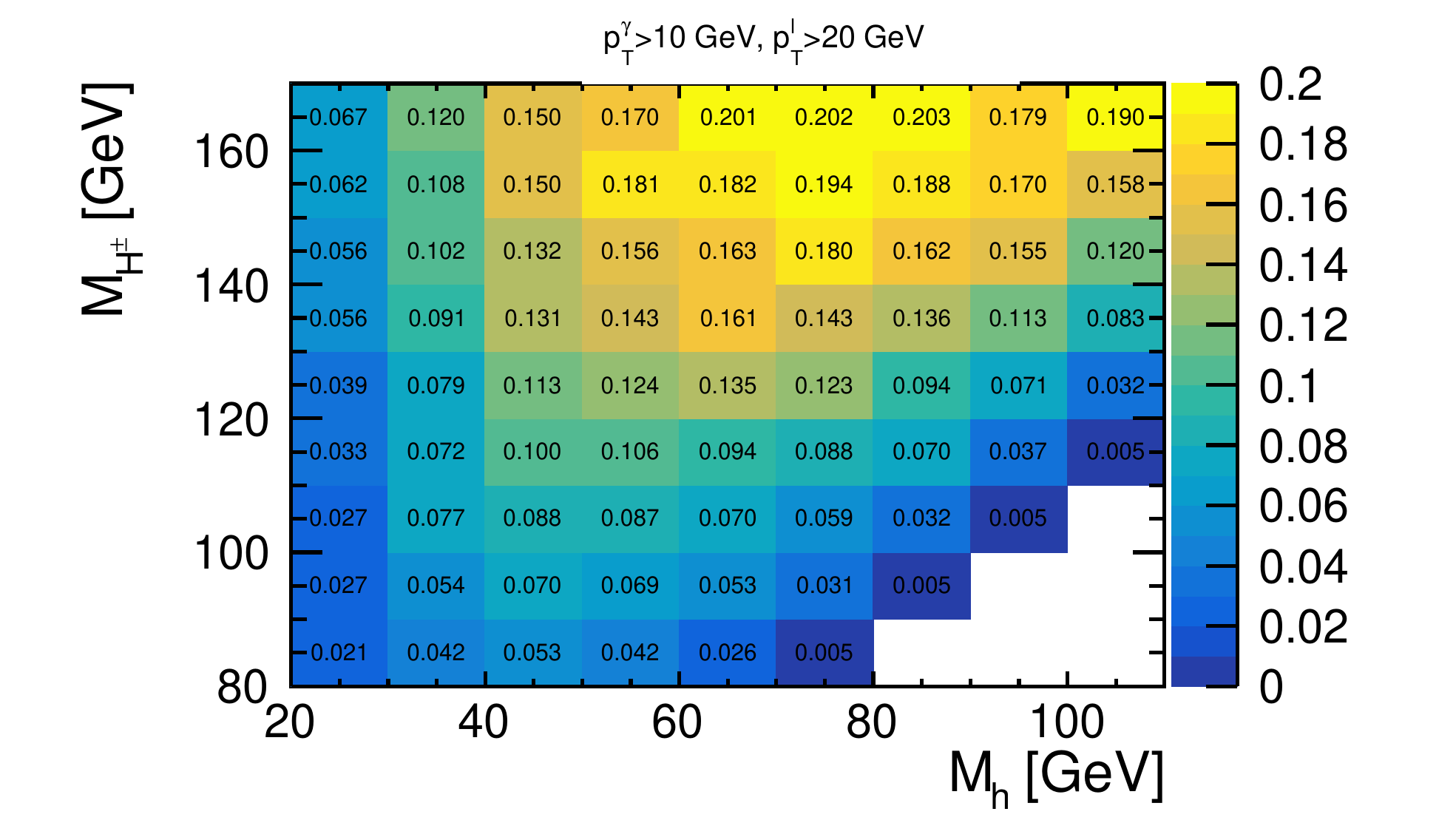} 
  \end{center}
 \end{minipage}
 \begin{minipage}{0.4\textwidth}
    \begin{center}    
 \includegraphics[height=4cm]{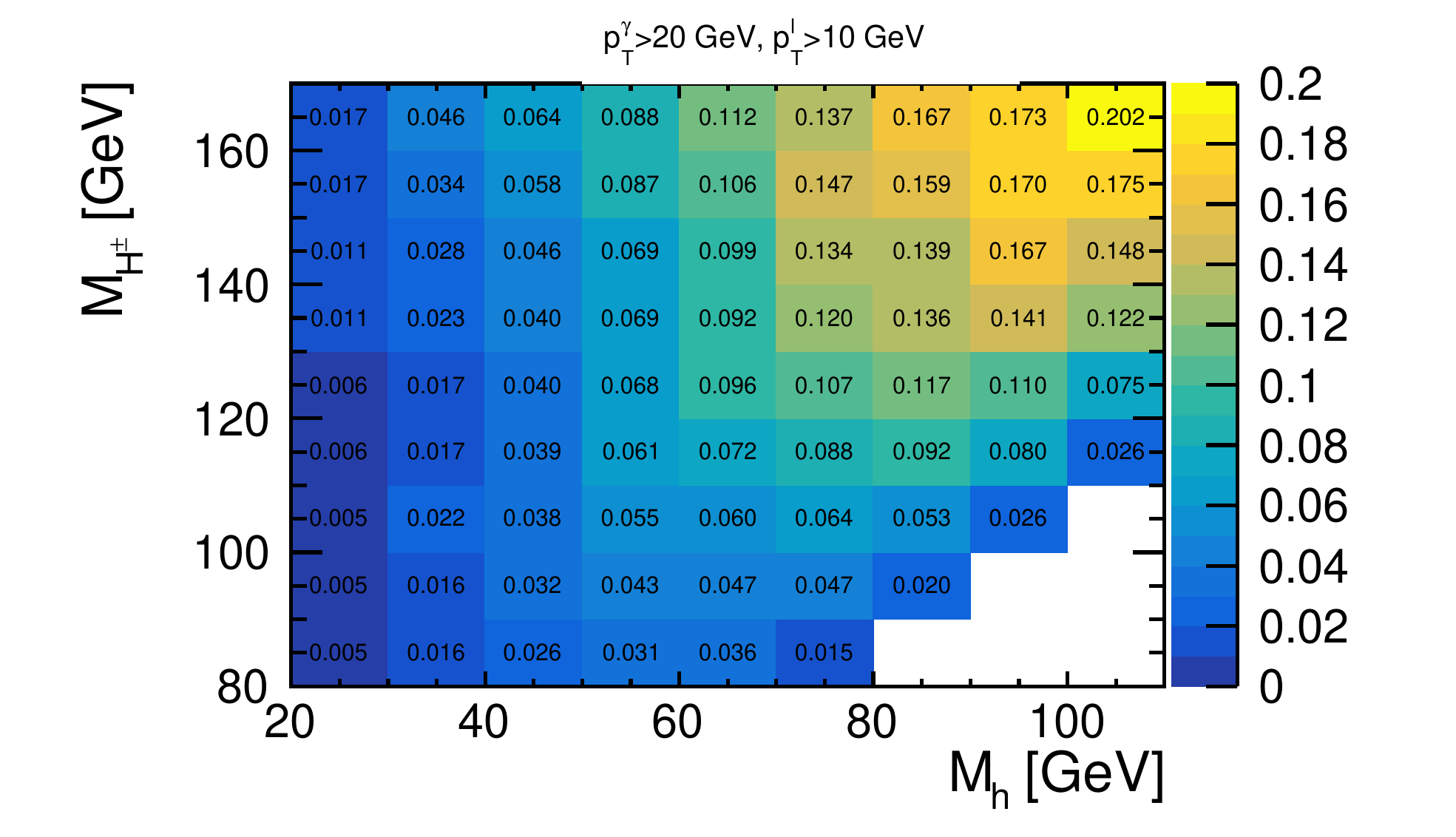} 
  \end{center}
 \end{minipage}
\begin{center}
\end{center}
 \begin{minipage}{0.4\textwidth}
    \begin{center}    
 \includegraphics[height=4cm]{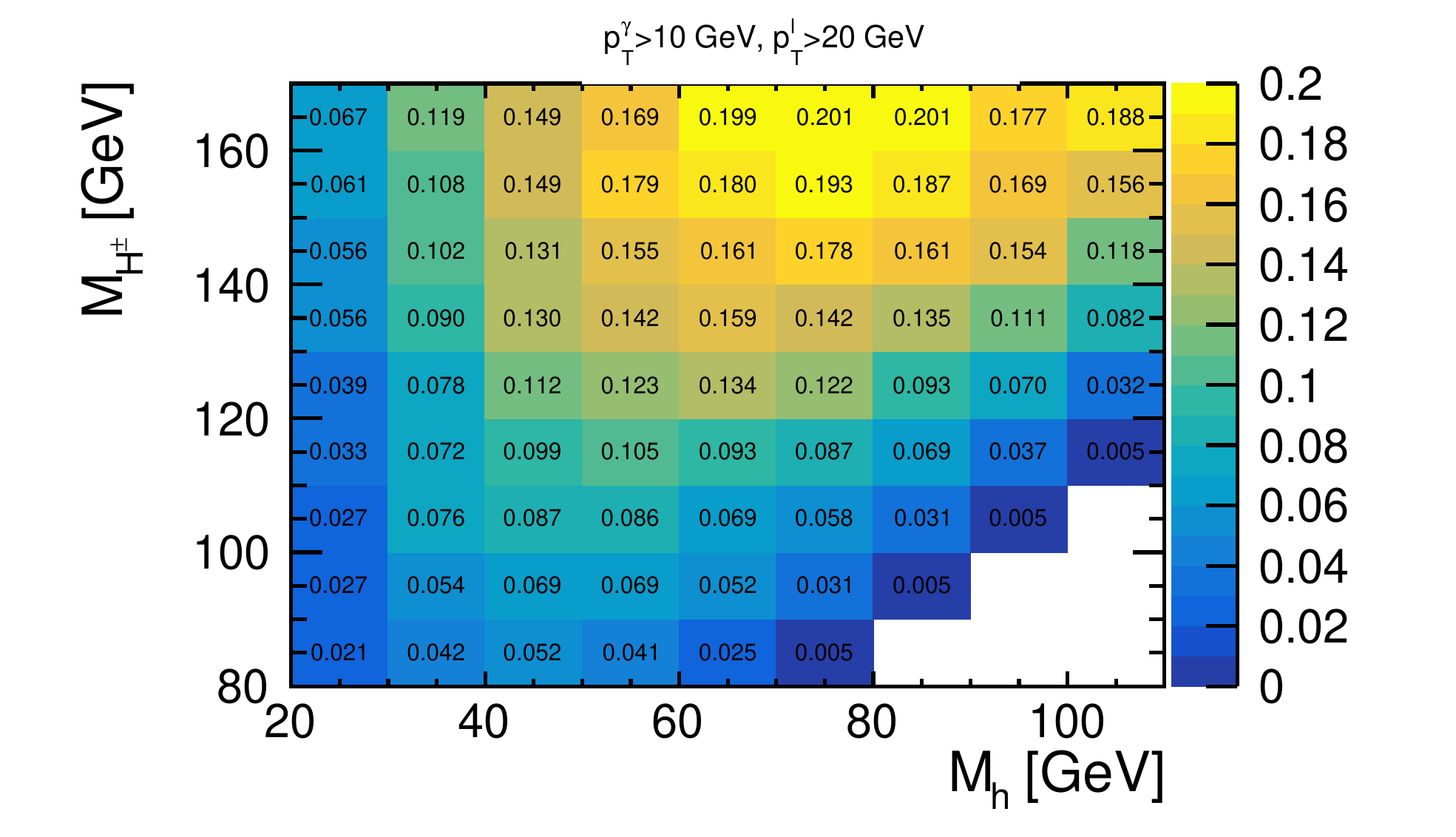} 
  \end{center}
 \end{minipage}
 \begin{minipage}{0.4\textwidth}
    \begin{center}    
 \includegraphics[height=4cm]{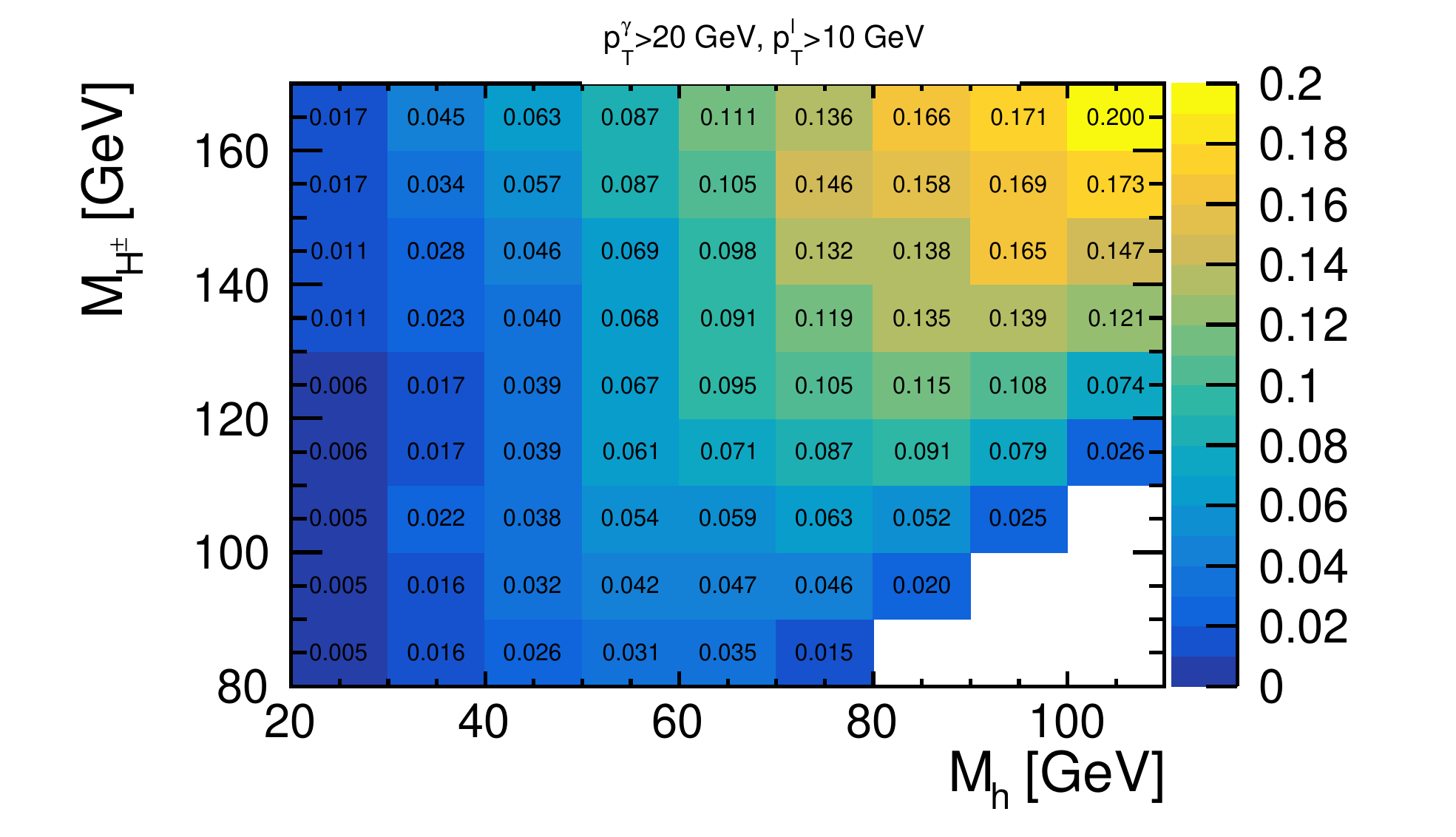} 
  \end{center}
 \end{minipage}
  \caption{Fiducial efficiency $\epsilon$ for detecting the signal via the  $\ell\nu_\ell+4\gamma$ signature at detector level for the two sets of cuts  provided, when $\sqrt{s}=13$ TeV (top) as well as  $\sqrt{s}=14$ TeV (bottom) and $L=$ 300 fb$^{-1}$.}\label{f_effi}
\end{figure}

By studying the acceptance efficiency $\epsilon_{\rm det}$  with the 14 BPs at the detector level, we have introduced two sets of cuts at the parton level: $p_{T}^{\gamma} > 10 ~\textrm{GeV}$ and $p_{T}^{\ell}>20 ~\textrm{GeV}$ plus $p_{T}^{\gamma} > 20 ~\textrm{GeV}$ and $p_{T}^{\ell}>10 ~\textrm{GeV}$. We provide the fiducial efficiencies for these two sets through the relation
\begin{equation}
\epsilon= \sigma(\textrm{cuts}) \times \epsilon_{\rm det} / \sigma(\textrm{no cuts}).
\end{equation}
We apply the efficiencies to the signal events and obtain the results shown in Fig~\ref{f_effi}.

\begin{figure}[t!]
\begin{center}
\end{center}
\begin{minipage}{0.4\textwidth}
    \begin{center}    
 \includegraphics[height=4cm]{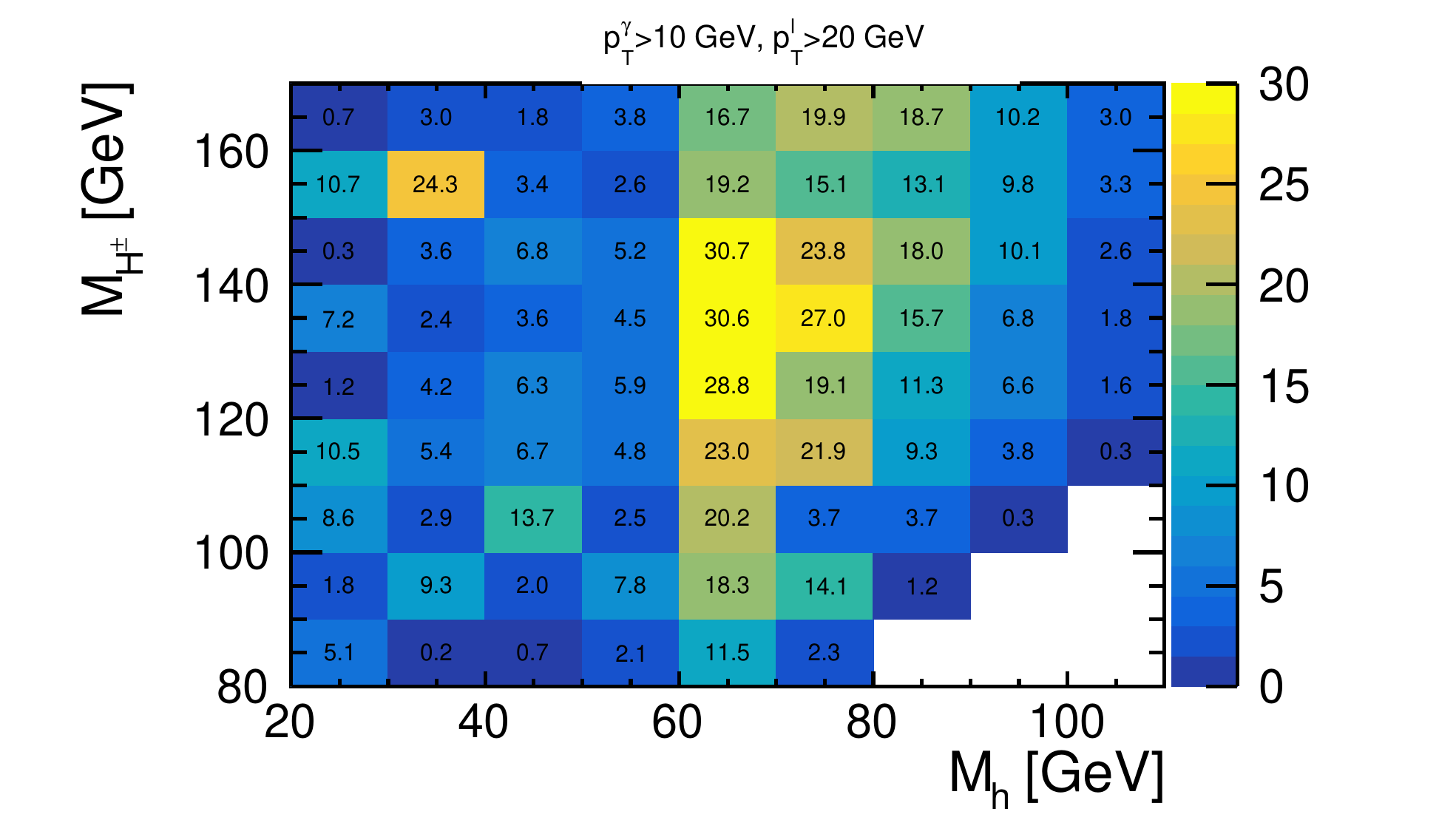} 
  \end{center}
 \end{minipage}
 \begin{minipage}{0.4\textwidth}
    \begin{center}    
 \includegraphics[height=4cm]{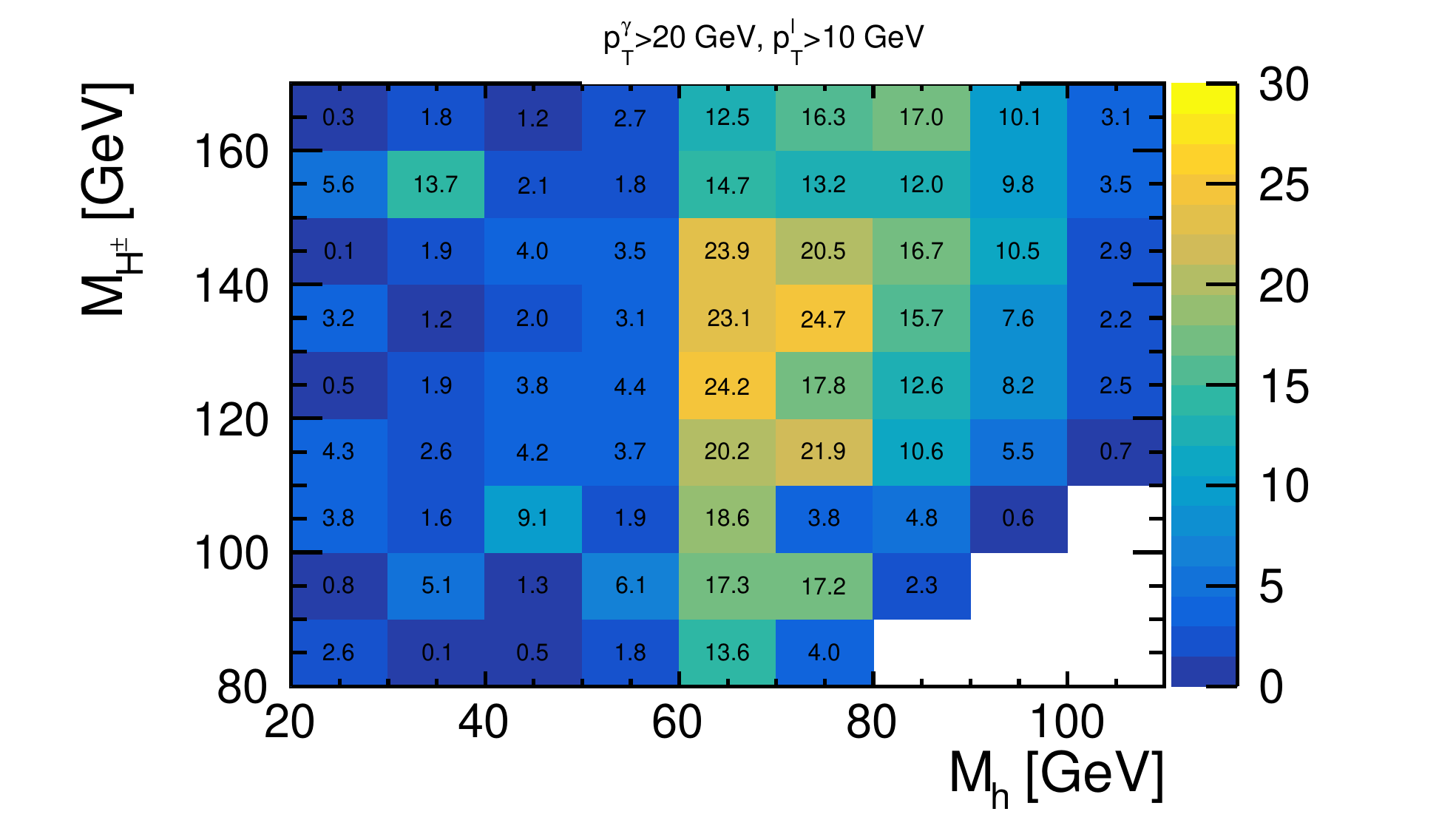} 
  \end{center}
 \end{minipage}
\begin{center}
\end{center}
 \begin{minipage}{0.4\textwidth}
    \begin{center}    
 \includegraphics[height=4cm]{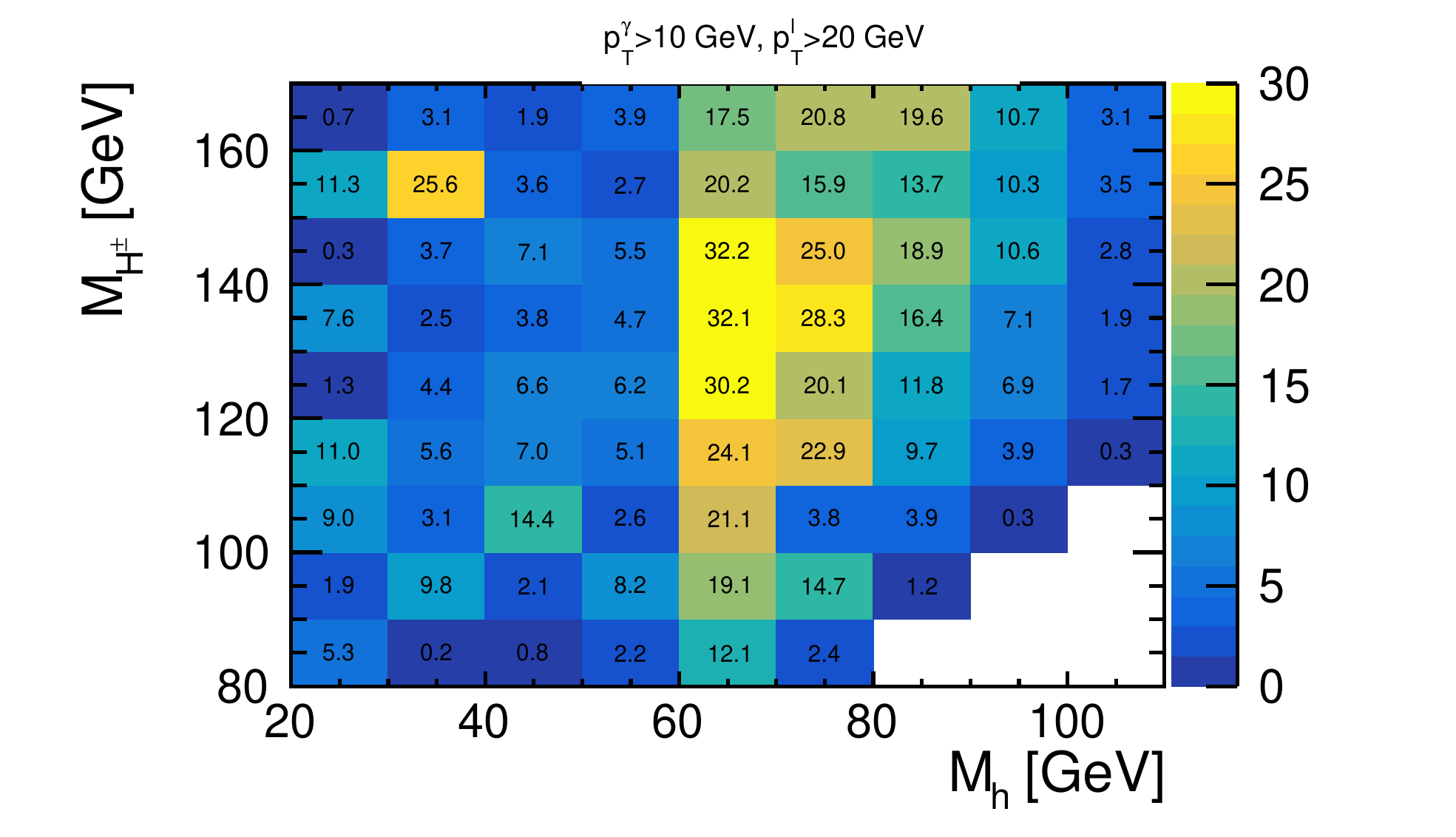} 
  \end{center}
 \end{minipage}
 \begin{minipage}{0.4\textwidth}
    \begin{center}    
 \includegraphics[height=4cm]{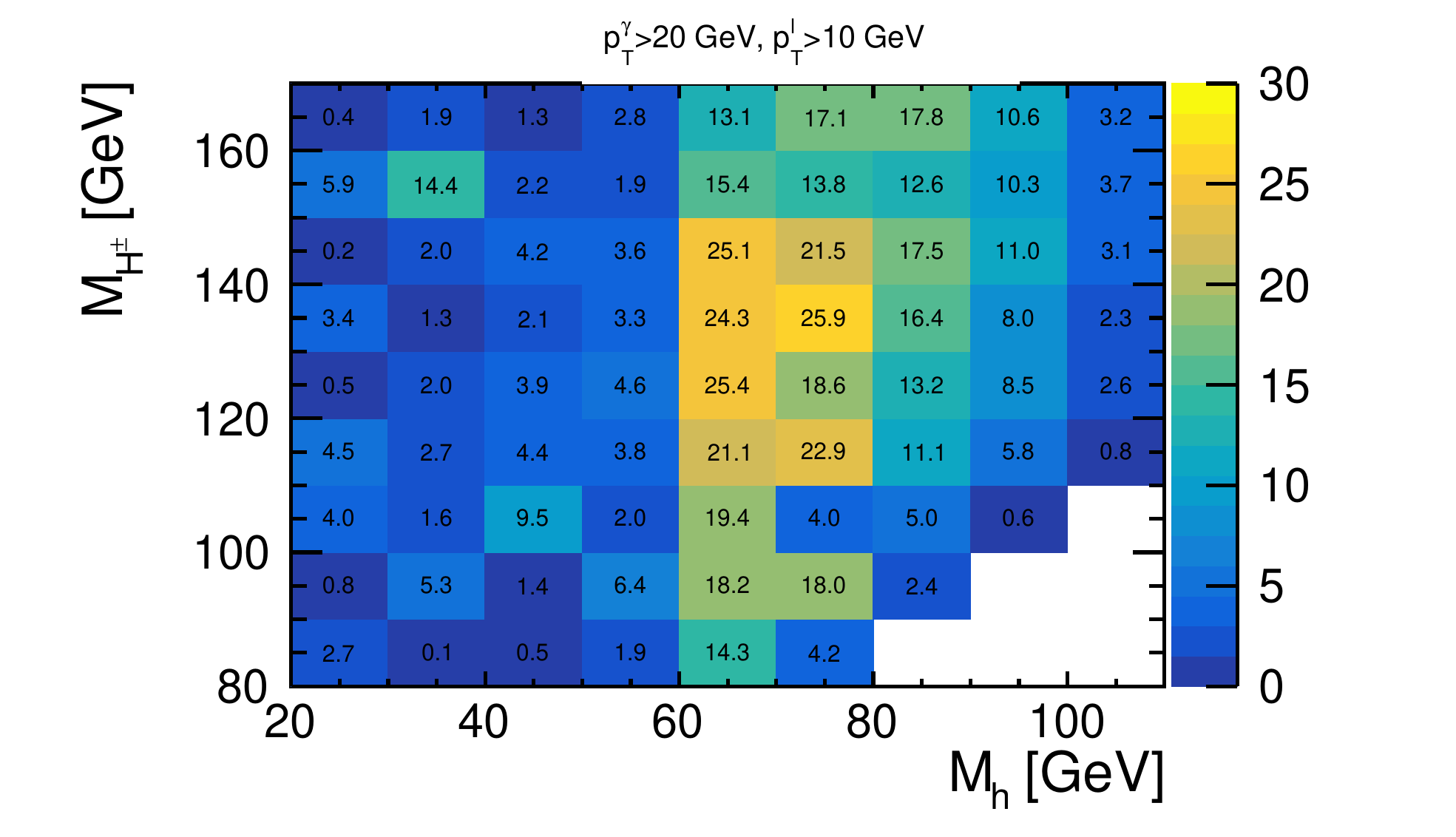} 
  \end{center}
 \end{minipage}
   \caption{The predicted significances over the  ($M_h, M_{H^\pm}$) plane for the two sets of cuts are shown,  when $\sqrt{s}=13$ TeV (top) as well as  $\sqrt{s}=14$ TeV (bottom) and $L=$ 300 fb$^{-1}$.}\label{f_sig_13}
\end{figure}

\begin{figure}[h!]
\begin{center}
\end{center}
\begin{minipage}{0.4\textwidth}
    \begin{center}    
 \includegraphics[height=4cm]{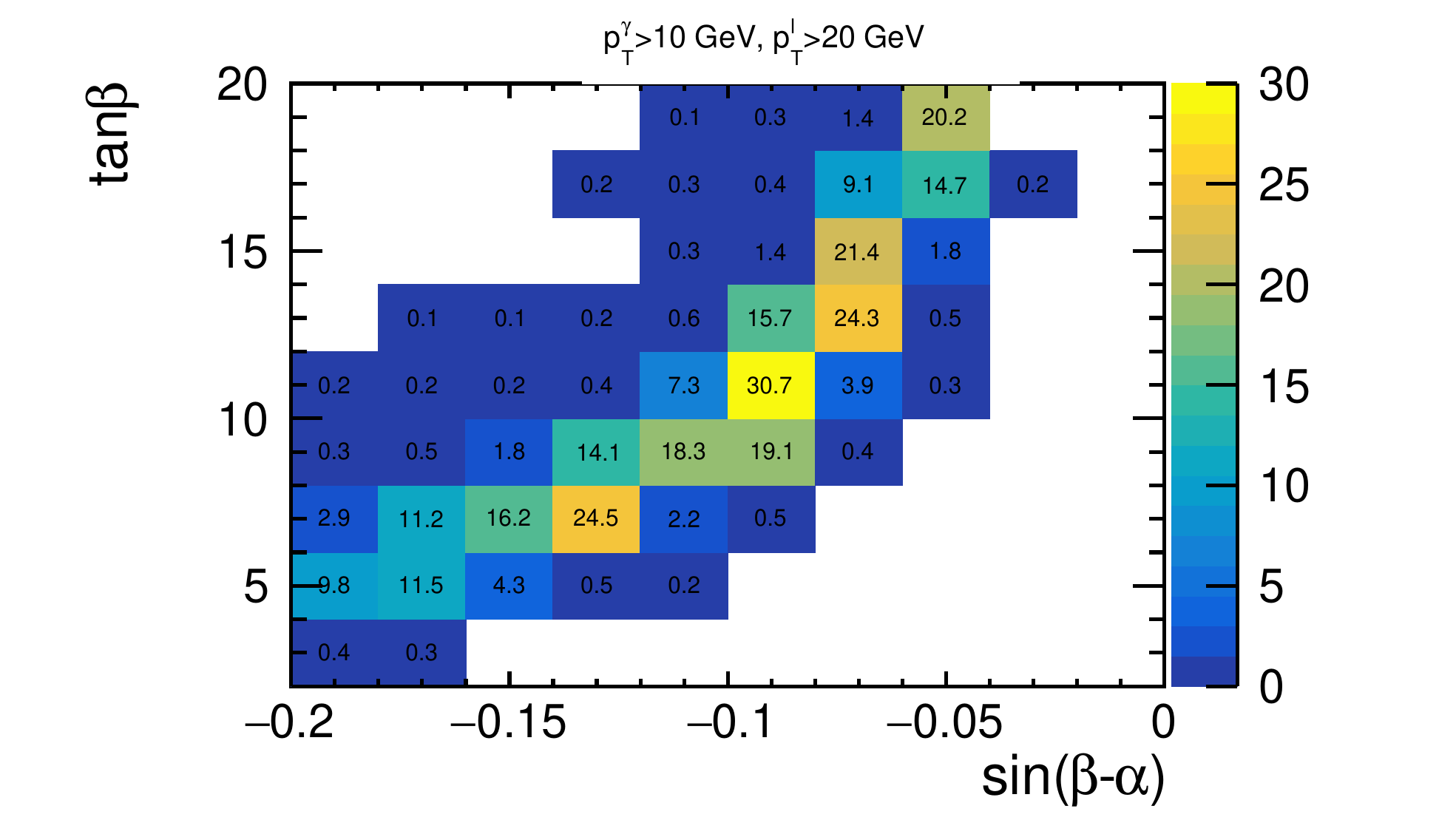} 
  \end{center}
 \end{minipage}
 \begin{minipage}{0.4\textwidth}
    \begin{center}    
 \includegraphics[height=4cm]{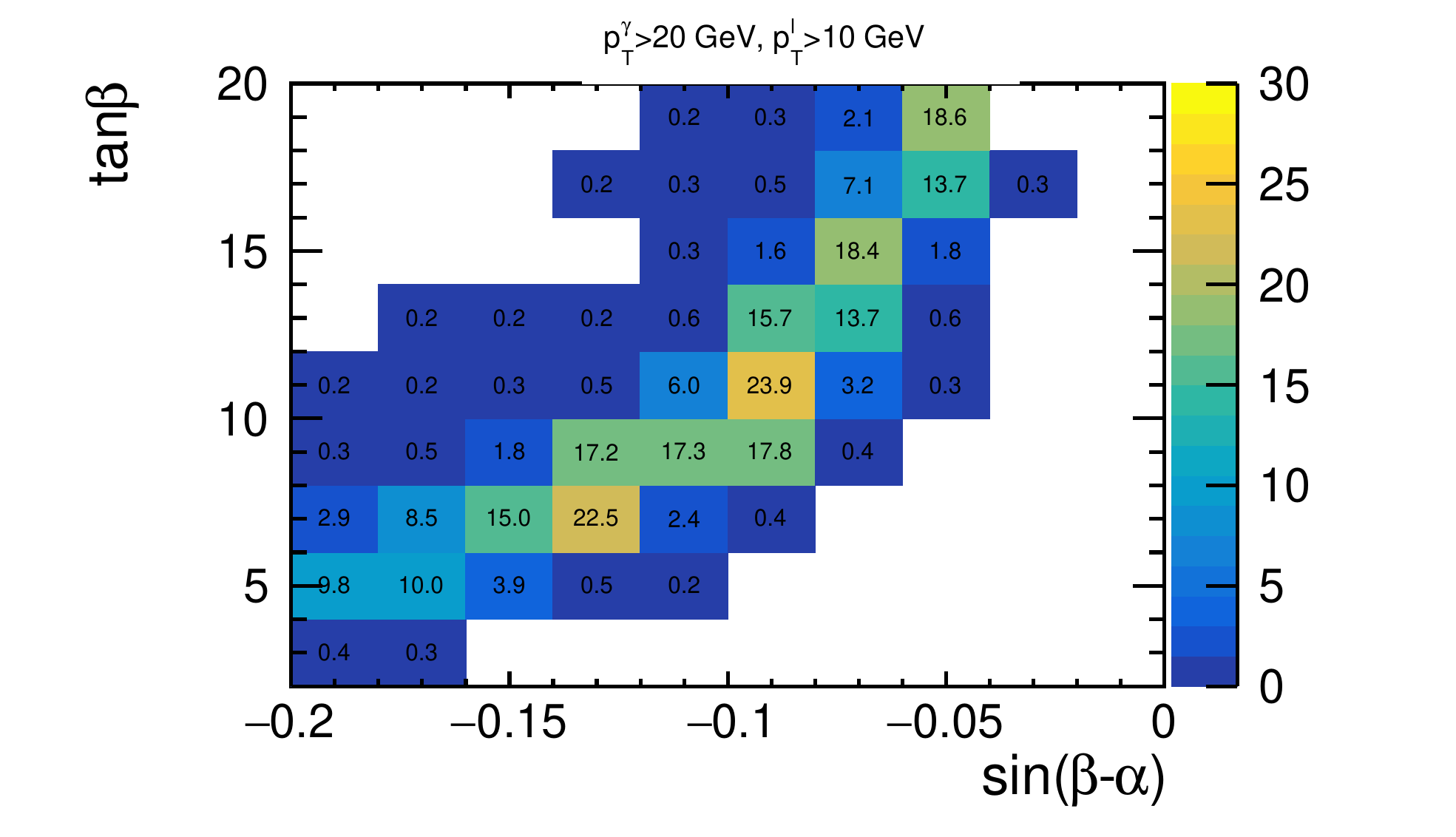} 
  \end{center}
 \end{minipage}
\begin{center}
\end{center}
 \begin{minipage}{0.4\textwidth}
    \begin{center}    
 \includegraphics[height=4cm]{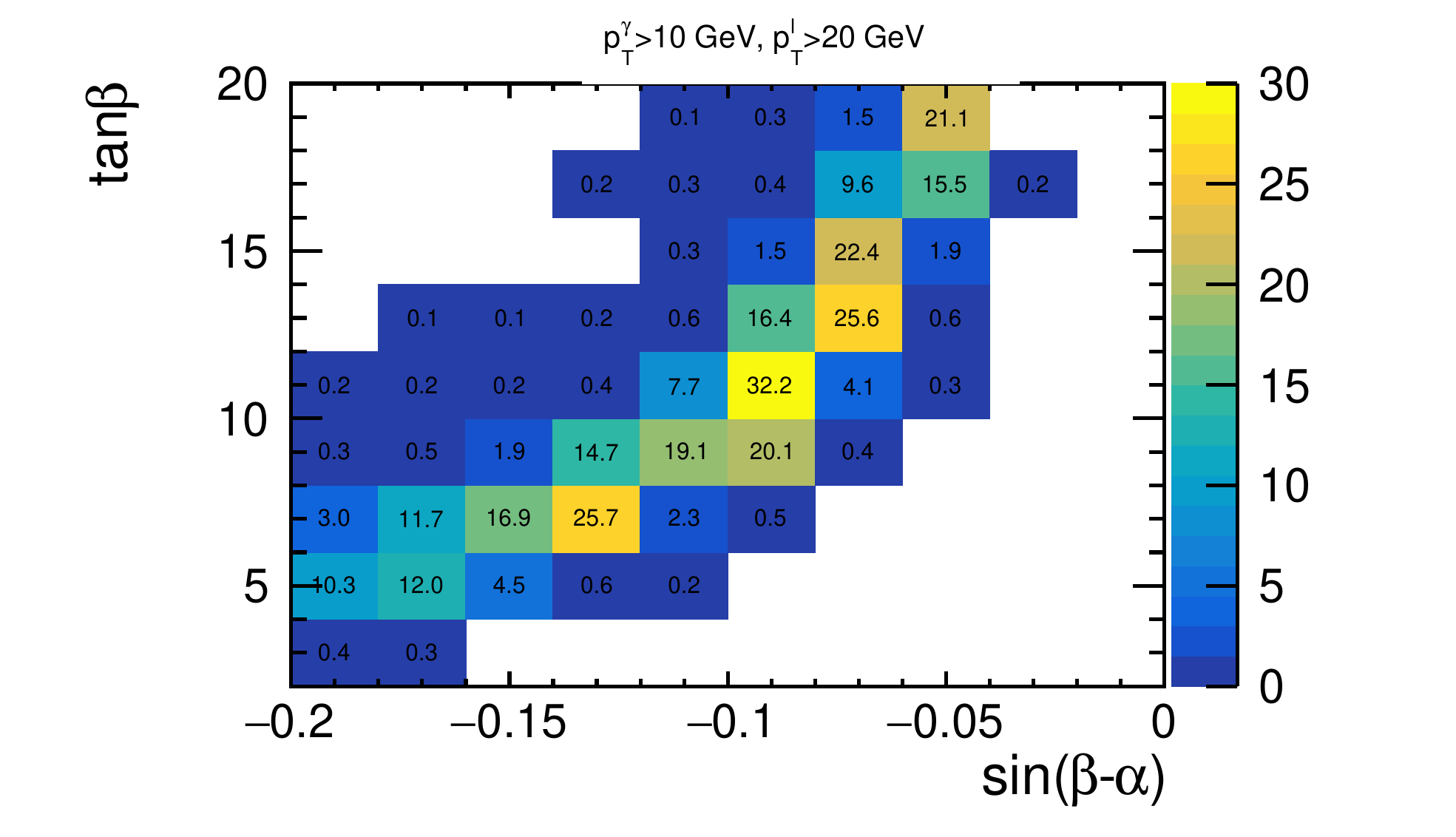} 
  \end{center}
 \end{minipage}
 \begin{minipage}{0.4\textwidth}
    \begin{center}    
 \includegraphics[height=4cm]{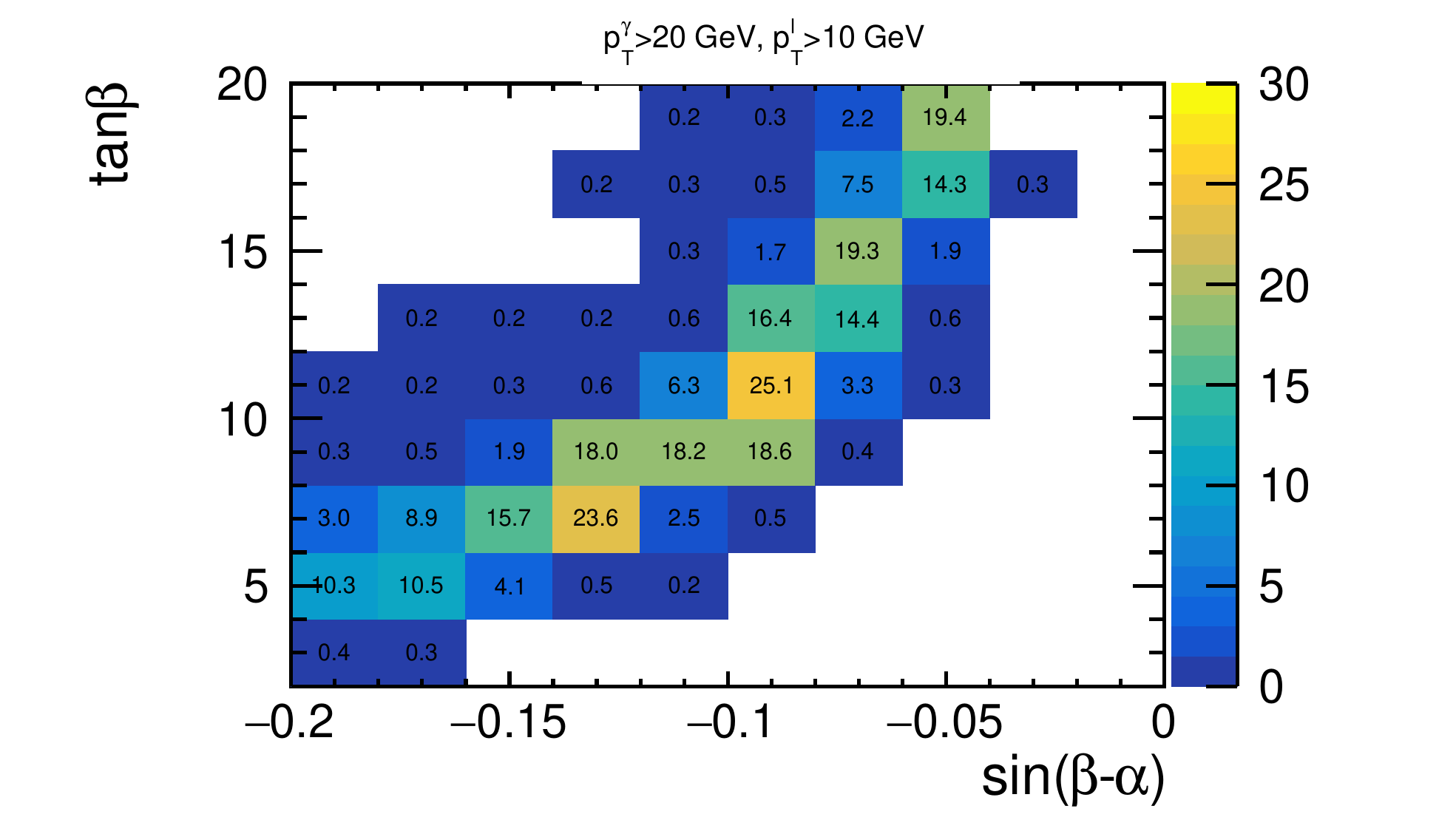} 
  \end{center}
 \end{minipage}
  \caption{The predicted significances over the ($\sin(\beta-\alpha), \tan\beta$)  plane for the two sets of cuts are shown, when $\sqrt{s}=13$ TeV (top) as well as  $\sqrt{s}=14$ TeV (bottom) and $L=$ 300 fb$^{-1}$.}\label{f_para_sig_13}
\end{figure}

The predicted significances for both energies and the given luminosity over the ($M_{h}, M_{H^{\pm}}$) plane are shown in Fig.~\ref{f_sig_13}. The significances are obtained by convoluting the signal production cross sections with cuts and acceptance efficiencies at detector level. For each point on the ($M_{h}, M_{H^{\pm}}$) plane, $tan\beta$ and $sin(\beta-\alpha)$ are allowed to vary in order to find the maximum significance.  
The predicted significances are then mapped over the $(sin(\beta-\alpha),tan\beta)$ plane, which are shown in Fig.~\ref{f_para_sig_13}.

In a summary, we have examined the feasibility of the signature $\wboson+4\gamma$, where the $W^\pm$ decays leptonically in electrons and muons, from the associated production of the charged Higgs boson and lightest neutral Higgs state in the Type-I scenario of the 2HDM (i.e., via $pp \to H^\pm h\to W^{\pm(*)}hh\to \ell\nu_\ell + 4\gamma$) at the LHC with a collision energy of $\sqrt{s}=13$ and $14~\text{TeV}$ and an integrated luminosity of $L= 300$ $\fbinv$. We have exploited a MC analysis at a detector level by including parton shower, hadronisation and heavy flavour decays. By doing so, we have
confirmed a previous study done at the parton level. As we have shown, even after taking into account background processes events generated by both real and fake photons (from jets), the signal is essentially  background free, so that significances only depend upon the signal cross sections and the integrated luminosity. We have also provided some reliable estimates for the detector efficiency and associated heat maps which can expedite an estimate of the signal significance over the relevant Type-I parameter space, which could be useful for LHC working groups. Finally, for more thorough experimental analyses, we have also chosen 14 BPs, where the 
$W^\pm$ boson can be either on-shell or off-shell, depending on the mass difference $M_{H^\pm}-M_{h}$. 
\vspace{6pt} 

\authorcontributions{All authors have contributed in equal parts to all aspect of this research.}

\funding{The work of AA, RB, MK and BM is supported by the Moroccan Ministry of Higher Education and Scientific Research MESRSFC and
CNRST Project PPR/2015/6. The work of SM is supported in part through the NExT Institute and the STFC Consolidated Grant No. ST/L000296/1.
Y. W. is supported by the `Scientific Research Funding Project for
Introduced High-level Talents' of the Inner Mongolia Normal
University Grant No. 2019YJRC001 and the scientific research
funding for introduced high-level talents of Inner Mongolia of China.  Q.-S. Yan's work is supported by the Natural Science Foundation of China Grant No. 11875260.}

\institutionalreview{Not applicable.}

\informedconsent{Not applicable.}

\dataavailability{Not applicable.}

\conflictsofinterest{The authors declare no conflict of interest.} 




\end{paracol}
\reftitle{References}
\bibliography{2hdmw4photon}
\end{document}